\documentclass[twocolumn]{biophys-new}
\usepackage[utf8]{inputenc}
\usepackage{graphicx}
\usepackage[colorlinks,allcolors=cyan!70!black]{hyperref}
\usepackage{caption}
\usepackage{subcaption}

\usepackage{lipsum}
\usepackage{amsmath}

\title{Poisson-Boltzmann based machine learning  (PBML) model for electrostatic analysis}
\runningtitle{Biophysical Journal Template} 

\author[1]{Jiahui Chen}
\author[2]{Yongjia Xu}
\author[3]{Xin Yang}
\author[4]{Zixuan Cang}
\author[3,*]{Weihua Geng}
\author[5,6,*]{Guo-Wei Wei}

\runningauthor{Chen et al.} 

\affil[1]{Department of Mathematics, University of Arkansas, Fayetteville, AR 72701, USA}
\affil[2]{Google LLC, 1600 Amphitheater Pkwy, Mountain View, CA 94043, USA}
\affil[3]{Department of Mathematics, Southern Methodist University, Dallas, TX 75275, USA}
\affil[4]{Department of Mathematics, North Carolina State University, Raleigh, NC 27695, USA}
\affil[5]{Department of Mathematics, Michigan State University, MI 48824, USA}
\affil[6]{Department of Biochemistry and Molecular Biology, Michigan State University, MI 48824, USA}

\corrauthor[*]{wgeng@smu.edu, wei@math.msu.edu}

\papertype{Article}

\begin{document}
\begin{frontmatter}

\begin{abstract}
Electrostatics is of paramount importance to chemistry, physics, biology, and medicine.
The Poisson-Boltzmann (PB) theory is a  primary model for electrostatic analysis. 
However, it is highly challenging to compute accurate PB electrostatic solvation free energies for macromolecules due to 
the nonlinearity, 
dielectric jumps,
charge singularity , 
and geometric complexity 
associated with the PB equation. 
The present work introduces 
a PB based machine learning (PBML) model for biomolecular electrostatic analysis. 
Trained with the second-order accurate MIBPB solver, 
the proposed  PBML model is found to be more accurate and faster 
than several eminent PB solvers in electrostatic analysis.  
The proposed PBML model can provide highly accurate 
PB electrostatic solvation free energy of new biomolecules or new conformations 
generated by molecular dynamics with much reduced computational cost.  
\end{abstract}

\begin{sigstatement}
This manuscript provides 
a Poisson-Boltzmann based machine learning (PBML) model for biomolecular electrostatic analysis. 
The features as the input to the ML models are generated with mathematical algorithms 
using biomolecular structures and force field. 
The learned model, which is trained using the most accurate PB solver MIBPB 
on more than 4000 biomolecules shows improved efficiency and accuracy 
in electrostatic analysis compared with the popular PB solvers.  

\end{sigstatement}

\end{frontmatter}

\section{Introduction}\label{Introduction}
Electrostatics is ubiquitous in the molecular world.  
The analysis of molecular electrostatics is of crucial importance 
to the bioscience research community. 
There are two significant types of electrostatic analyses, namely,  
{\it qualitative} analysis for general electrostatic characteristics, 
such as visualization and electrostatic steering, 
and {\it quantitative} analysis for statistical, thermodynamic and/or kinetic observable, 
such as solvation free energy, solubility, and partition coefficient.   

Molecular electrostatics can be analyzed by explicit or implicit models. 
Explicit solvent models  
resolve electrostatic effect in atomic detail and thus are more accurate 
but can be very expensive for large biomolecular systems. 
Implicit solvent models describe the solvent as a dielectric continuum, 
while the solute molecule is modeled with an atomistic description  
\cite{Honig:1995a}.
A wide variety of  two-scale implicit solvent models 
has been  developed for  electrostatic analysis, 
including  generalized Born (GB)  
\cite{Onufriev:2000},
polarizable continuum  
\cite{Tomasi:2005}
and Poisson-Boltzmann (PB) models 
\cite{Fogolari:2002}.

%
PB models have been applied to  calculating  protein titration states
\cite{Bashford:1990}, 
protein-protein and protein-ligand binding energetics 
\cite{
Onufriev:2013}, 
RNA nucleotide protonation\cite{Tang:2007},
chromatin packing \cite{Zhang:2003a}, etc.
The PB theory has also been used for the evaluation of biomolecular electrostatic forces for molecular Langevin dynamics 
or Brownian dynamics \cite{UHBD}.  GB methods are faster than PB methods,
but provide only heuristic estimates for PB electrostatic energies. 
%
%

Due to its success in describing biomolecular systems, 
the PB model has attracted a wide attention in both mathematical and biophysical communities.  
In the past two decades, many efforts have been given to the development of  accurate, efficient, reliable  and  robust PB solvers.   
A large number of methods have been proposed in the literature, 
including the finite difference method (FDM) \cite{Jo:2008}, 
finite element method (FEM) \cite{Baker:2001b}, 
and boundary element method (BEM) \cite{Geng2013:3,LuBenzhuo:2013}. 
Among them, the FDM is prevalently used in the field due to its simplicity in implementation. 
The emblematic solvers in this category are 
Amber PBSA \cite{PBSA:2009}, 
Delphi \cite{Delphi:2012}
APBS 
\cite{Baker:2001}, 
MIBPB \cite{DuanChen:2011a}, 
CHARMM PBEQ \cite{Jo:2008}, 
etc.  

The PB equation is a nonlinear elliptic equation 
with singular source term. 
For biomolecules, the continuum-discrete interface is nonsmooth, 
resulting in a nonlinear elliptic interface problem 
with discontinuous coefficients and singular source terms \cite{geng2017two}. 
These difficulties make finding 
the numerical solution to the PB equation for biomolecules challenging.
There are numerous efforts  in  developing high order methods for elliptic interface problems 
in the past two decades \cite{Peskin:1977,LeVeque:1994, Yu:2007c}. 
Among them, matched interface and boundary (MIB) method 
offers  arbitrarily  high-order  accuracy  in  principle  
and  up  to  sixth-order  accurate  MIB schemes 
have been constructed for three-dimensional complex interfaces \cite{Yu:2007c}. 
The MIB based second-order accurate PB solver, the MIBPB, was constructed to take care of biomolecular interfaces 
and singular charges \cite{Geng:2007a}. 
The recent development of the Eulerian solvent excluded surface (ESES) \cite{ESES:2017}, 
which provides analytical biomolecular surface representation in the Cartesian domain,  
improves the stability and robustness of the MIBPB solver.
 
Nonetheless, the generation of highly accurate electrostatic potentials for large biomolecules can be extremely expensive. 
For example, it takes days to solve the PB model on
a protein with about 50,000 atoms at the mesh of 0.2 \AA~ on a single CPU. 
Additionally, the information generated for the electrostatic analysis of a given biomolecule is not 
transferable 
to other proteins. 
Therefore, one has to carry out 
the separated electrostatic analysis of different proteins or the same protein with different protonation states or conformations.
These issues call for innovative approaches, 
such as machine learning and dynamic programming, 
to biomolecular electrostatic analysis.   

Recently we have witnessed the explosion of machine learning studies 
in science and engineering. 
In particular, deep neural networks which 
discover intricate structures in large datasets, 
have fueled the rapid growth 
in application such as 
computer vision,  
natural language processing,  
speech  recognition, 
handwriting recognition,  
\cite{
lecun2015deep, 
korotcov2017comparison,jimenez2018k}, etc.
Machine learning has become an indispensable tool 
in the analysis and prediction of  large and diverse molecular and biomolecular data sets,  
including  bioactivity of small molecular drugs \cite{hughes2015modeling} 
%
and genomics \cite{lam2018hybrid}. 
Studies in computational biology and biophysics, 
such as the predictions of solvation free energies,  
protein-ligand  binding affinities, 
mutation impacts, 
toxicity, 
partition coefficients,  
B-factors etc. 
adopt machine learning approaches 
\cite{
sunseri2016d3r, 
ZXCang:2018a,
KDWu:2018a, 
KDWu:2018b}.
These developments open the door for machine learning based electrostatic analysis.  

The objective of the present work is to develop a machine learning solution of the PB equation 
for  the electrostatic analysis of 
biomolecules. 
To this end, we first construct an accurate and efficient mathematical representation of electrostatic potential 
to effectively characterize the probability distribution of 
biomolecular 
electrostatics. 
Theoretically, the exact form of this distribution is not available 
even if  one solves the PB equation for all possible 
biomolecular 
structures. 
However, in practice, this probability distribution can be sampled by using a PB solver, 
which provides machine learning training labels. 
Our approach is based on a {\it representability} hypothesis 
and a {\it learning} hypothesis.  
The {\it representability} hypothesis states that the  electrostatic potential 
of a biomolecule 
can be described by a set of partial charges 
and their geometric relations to the solvent. 
This hypothesis guilds the construction of the feature vector for the characterization of  the probability distribution of biomolecular 
electrostatics. 
The {\it learning} hypothesis states that  
biomolecular 
electrostatics can be effectively represented by a feature vector as described by the representability hypothesis. 
When the probability distribution of biomolecular 
electrostatics 
is sufficiently sampled from a training set, 
a machine learning model can be established  based on training labels and associated feature vectors 
to accurately predict the electrostatic potential of an unseen dataset 
which shares the same probability distribution with the training set. 

The protocol described above calls for an accurate PB solver, 
which calculates machine  learning labels and thus the probability distribution of molecular and biomolecular electrostatics. 
To this end, 
we apply the accurate MIBPB solver \cite{Geng:2007a} at a refined mesh size 0.2 \AA~to generate solvation energy labels 
to minimize the numerical errors.  
 
The representability hypothesis does not specify how to construct an accurate and efficient representation.  
An average biomolecule in the human body consists of about 6,000 atoms 
that lie in an $18,000$-dimensional Euclidean space (${\mathbb R}^{18,000}$). 
Such a high-dimensionality makes the first principle calculations intractable. 
Additionally, using macromolecular structures in deep convolutional neural networks (CNNs)
is extremely expensive. 
For example, the 3D coordinates representation of a biomolecule 
with about 50\AA~ side length at a low resolution of $0.5$\AA~ 
requires feature dimension of $100^{3n}$, where $n$ is the number of element types. 
The variable sizes of biomolecules also hinder the application of machine learning algorithms. 
These challenges motivate the development  of scalable 
and intrinsically low-dimensional representations of biomolecular structures.  
Our hypothesis is that intrinsic physics lie in low-dimensional manifolds or spaces 
embedded in a high dimensional data space \cite{ZXCang:2018a}.
Recently, a few low-dimensional representations for 
biomolecules 
have been developed in terms of  algebraic topology 
\cite{KLXia:2014c, ZXCang:2018a},
%
differential geometry 
\cite{DDNguyen:2016c},
and graph theory 
\cite{KLXia:2013d}.
 All of these approaches can be used to represent biomolecular electrostatics. 
In this work, we adopt the graph theory representation due to its simplicity, 
in conjuration with a collection of features from the fast GB models to predict electrostatics from the PB model. 

The rest of this paper is organized as follows. 
After this Introduction section, the Materials and Methods section describes models and algorithms used in the present work. 
We give a brief review of the PB model, the GB model, the graph theory, and machine learning algorithms used in developing the proposed Poisson-Boltzmann based machine learning (PBML) model for biomolecular electrostatic analysis. 
Simulation results and related discussion are presented in the Results section. 
Various  convergence tests have been carried out to search for the most accurate PB solver 
to calculate  machine learning labels. 
Our feature vectors are optimized with respect to from a few simple machine learning algorithms, namely, linear regression, random forest, and gradient boosted decision trees, to more complicated deep neural network (DNN). 
We demonstrate that the proposed PBML  model is more accurate and reliable than commonly used PB solvers. 
This paper ends with a Conclusion section. 
 
\section*{Materials and Methods}

\label{TheoryMethods}

In  this section, we briefly review essential concepts and methods underpinning the proposed PBML model.

\subsection*{The Poisson-Boltzmann (PB) model }
\begin{figure}[htb]
\setlength{\unitlength}{1cm}
\begin{center}
\begin{tabular}{llll}
\raisebox{83pt}{(a) }&\hskip -12pt  \includegraphics[width=1.3in]{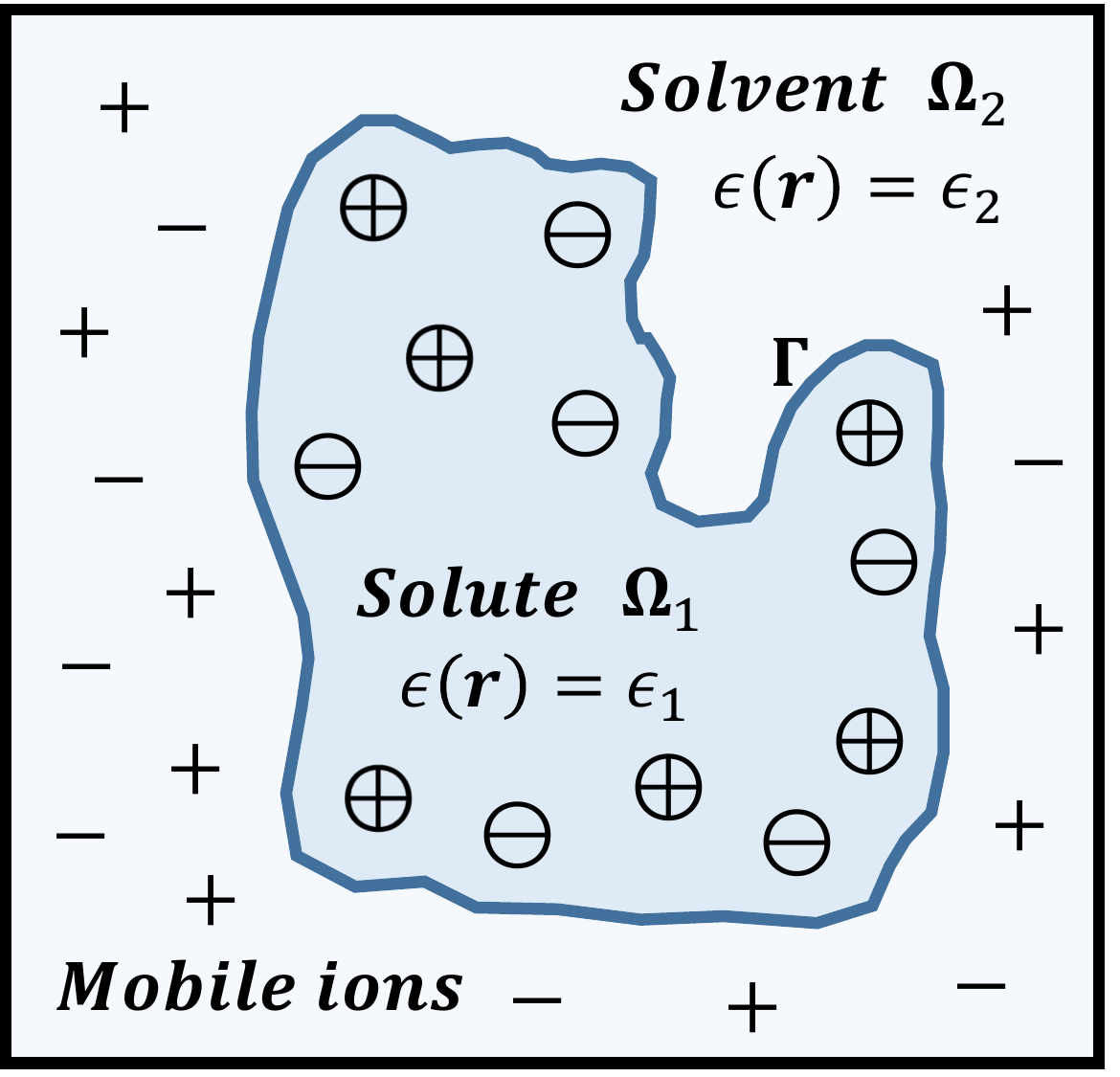}
\raisebox{83pt}{(b) }&\hskip -12pt \includegraphics[width=1.3in]{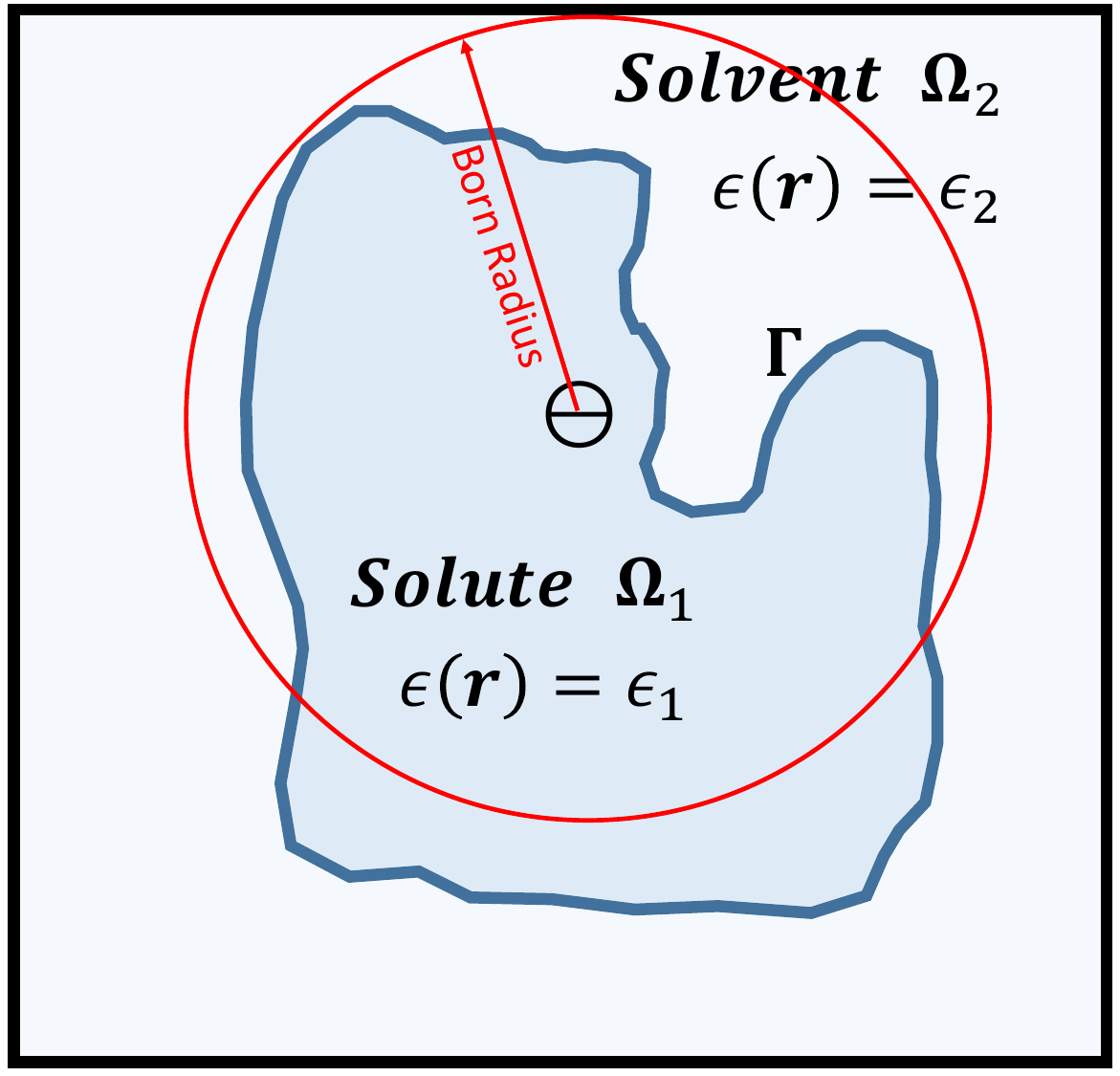}
\end{tabular}
\caption{\small An illustration of  the PB model and the GB model. (a): the PB model with solute region $\Omega_1$ and  solvent region $\Omega_2$, separated by molecular surface $\Gamma$, (b): the GB model represented by spherical cavities with Born radii and centered charges (one is shown here).}
\label{fig_1}
\end{center}
\end{figure}

As shown in Fig. \ref{fig_1}(a), the PB model governs 
electrostatics with the interior solute domain $\Omega_1$ with fixed charges $q_k$ 
located at atomic centers ${ \bf r}_k$ for $k=1,...,N_c$,
and the exterior solvent domain $\Omega_2$ with dissolved ions approximated by the Boltzmann distribution.
These two domains are separated by the dielectric interface $\Gamma$. 
Among a variety of surface models, the most commonly used one is the solvent excluded surface or molecular surface \cite{SESConnolly:1983}.
For simplicity, a linearized PB model is considered in the present work as
\begin{equation}
-\nabla \cdot \epsilon({\bf r}) \nabla \phi({\bf r}) + \bar{\kappa}^2({\bf r})\phi({\bf r}) =
\sum_{k=1}^{N_c} q_k \delta({\bf r}-{\bf r}_k),
\label{eqNPBE}
\end{equation}
where  $\phi(\bf{r})$ is the electrostatic potential and $\epsilon(\bf{r})$ are the dielectric constants given by 
\begin{equation}
\epsilon(\bf{r})=\left\{
\begin{array}{ll}
\epsilon_1, \quad \bf{r} \in \Omega_1,
\\
\epsilon_2, \quad \bf{r} \in \Omega_2,
\end{array}
\right.
\label{eqEpsilon}
\end{equation}
and $\bar{\kappa}$ is the screening parameter with the relation $\bar{\kappa}^2 = \epsilon_2 \kappa^2$ where $\kappa$ is the inverse Debye length measuring the ionic effective length.
The PB model has the interface conditions on the molecular surface defined as
\begin{equation}
\phi_1(\bf{r}) = \phi_2(\bf{r}),
\quad
\epsilon_1 \frac{\partial\phi_1(\bf{r})}{\partial \bf{n}}=\epsilon_2 \frac{\partial \phi_2 (\bf{r})}{\partial \bf{n}},
\quad
\bf{r} \in \Gamma
\label{eqInterface}
\end{equation}
where $\phi_1$ and $\phi_2$ are the limit values when approaching the interface from the inside or outside the solute domain, and $\bf{n}$ is the outward unit normal vector on $\Gamma$. 
The lack of appropriate or rigorous treatments of these interface conditions is the major error source for   many existing  PB solvers.  
The  far-field boundary condition for the PB model is $\lim_{|\bf{r}|\rightarrow\infty}\phi(\bf{r}) = 0$, 
which is approximated using the Screened Coulombic potential.

The PB electrostatic solvation free energy is obtained by 
\begin{equation}
 \Delta G= \frac{1}{2}\sum_{k=1}^{N_c}q_k (\phi({\bf r}_k) -\phi_0({\bf r}_k) )
\label{solvationEnergy}
\end{equation}
where $\phi_0({\bf r}_k)$ is the solution of the PB equation as if there were no solvent-solute interface. 
Note that near the interface $\Gamma$, 
the interpolation of $\phi({\bf r}_k)$ in Eq. (\ref{solvationEnergy}) using $\phi$ at the grid points can be another major error source.  
An interface based scheme  like MIB is required to interpolate $\phi({\bf r}_k)$ \cite{DDNguyen:2017a}. 

\subsection*{The generalized Born (GB) model}
With the same mathematical setting  as that  for the PB model, the Generalized Born (GB) model is devised to approximate the PB model. Compared with the PB model, the GB model offers a relatively simple and computationally more efficient approach to compute the long-range electrostatic interactions in biomolecules, which is the bottleneck in classical all-atom simulations. 
As illustrated in Fig.~\ref{fig_1}(b), the GB approximation of electrostatic solvation free energy can be expressed as the superposition of spherical cavities with effective Born radii and centered charges (only one is shown in the figure) \cite{forouzesh2017grid} : 
\begin{flalign}
\label{GB_eqn}
\Delta G^{\rm GB} & \approx \sum_{ij}\Delta G^{\rm GB}_{ij} \\
& = \frac{1}{2} \Big( \frac{1}{\epsilon_2}-\frac{1}{\epsilon_1} \Big) \frac{1}{1+\alpha\beta} \sum_{ij} q_i q_j \Big( \frac{1}{f_{ij}(r_{ij}, R_{i}, R_j)} + \frac{\alpha\beta}{A} \Big) \nonumber
\end{flalign}
where $r_{ij}$  the distance between atoms $i$ and $j$, 
$\beta = \epsilon_1/\epsilon_2$, 
$\alpha = 0.571412$, 
$A$ the electrostatic size of the molecule,  
using reciprocal of the Born radius $R^{-1}_i = {\Big( -\frac{1}{4\pi} \oint_{\Gamma } \frac{{\bf r}-{\bf r}_i}{|{\bf r}-{\bf r}_i|^6} \cdot \text{d}\bf{S} \Big)}^{1/3}$,
\begin{equation}
\label{f_ij}
f_{ij} = \sqrt{r^2_{ij}+R_iR_j {\rm exp}\Big( -\frac{r^2_{ij}}{4R_iR_j} \Big)},~~
\end{equation}
To carry out the boundary integral in evaluating $R^{-1}_i$, 
the MSMS package \cite{Sanner:1996} is used for the triangulation of $\Gamma$. 
Note this integration is the most time-consuming step in the GB calculation. 
The ESES softwware \cite{ESES:2017} can be used to improve the current GB model if a higher level accuracy is desired. 
 
\subsection*{The graph theory representation}\label{sec:mwcg}

 Graph theory is a prime subject of discrete mathematics and concerns graphs as mathematical structures for modeling pairwise relations between vertices, nodes, or points. Such pairwise relations define graph edges. Algebraic graph theory, particularly spectral graph theory, studies algebraic connectivity, characteristic polynomial, and eigenvalues and eigenvectors of matrices associated with the graph, such as adjacency matrix or Laplacian matrix. 
Graphs have been widely  used in chemistry 
and biomolecular modeling 
\cite{Bahar:1997}.
However, the diagonalization of the interaction Laplacian matrix has the computational complexity of  ${\cal O}(N^3)$ with $N$ being the number of matrix elements. Alternatively, geometric graph theory bypasses the time-consuming matrix diagonalization and can be made of  ${\cal O}(N)$ in computational complexity \cite{KLXia:2013d}.

In conjugation with machine learning algorithms,  multiscale weighted colored subgraph (MWCS)  was found to outperform many other methods in representing complex biomolecular structures \cite{bramer2018multiscale,DDNguyen:2018e}. We first consider weighted colored subgraph (WCS) to describe  electrostatic interactions in a protein of $N$ atoms.
It incorporates  kernels to characterize pairwise distance-weighted atomic correlations. All interactions are classified  according to element types, leading to colored subgraphs.   To  use WCS for analyzing  protein electrostatic interactions, we formulate  all the atoms and their pairwise interactions  into a  weighted graph $G(V, E)$ with vertices $V$ and edges $E$. As such, the $i$th atom is labeled by both its position ${\bf r}_i$  and element type $\alpha_i$. Therefore, we express vertices $V$ as
\begin{align}
V=\{({\bf r}_i,\alpha_i)| {\bf r}_i\in \mathbb{R}^3, \alpha_i\in \mathcal{C}, i=1,2,\dots, N\},
\end{align}
  where $\mathcal{C}=\{{\rm  C, N, O, S, H} \}$ contains all the commonly occurring element types in a protein. Obviously, for different biomolecular systems, we need to modify  $\mathcal{C}$ accordingly.

To describe pairwise interactions between atoms in a  protein,  we define a colored set  $\mathcal{P}=\{\alpha \beta\}$ with $\alpha, \beta \in \mathcal{C}$. 
For each subset of element pairs $\mathcal{P}_k$, $k=1,2,\dots,15$, a set of involved vertices $V_{\mathcal{P}_k}$ is a subset of $V$ containing all atoms that belong to the pair in $\mathcal{P}_k$. For example, a partition $\mathcal{P}_2 = \{{\rm  CN}\}$ contains all pairs of atoms  in the protein with one atom being a  carbon and another atom being a nitrogen.  Based on this setting, all the edges in such WCS describing pairwise atomic interactions are defined by
\begin{flalign}
		E_{\mathcal{P}_k}^{\sigma,\tau,\zeta} = \{ \Phi^\sigma_{\tau,\zeta}(\|{\bf r}_i - {\bf r}_j\|)|  \alpha_i\beta_j \in\mathcal{P}_k; i,j=1,\dots,N\}
\end{flalign}
where $\|{\bf r}_i - {\bf r}_j\|$ defines a Euclidean distance between $i^{th}$ and $j^{th}$ atoms, $\sigma$ indicates the type of radial basic functions (e.g., $\sigma=\text{L}$ for Lorentz kernel, $\sigma=\text{E}$ for exponential kernel), $\tau$ is a scale distance factor between two atoms, and $\zeta$ is a parameter of power in the kernel (i.e., $\zeta=\kappa$ when $\sigma=\text{E}$, $\zeta=\nu$ when $\sigma=\text{L}$). The kernel  $\Phi^\sigma_{\tau,\zeta}$ characterizes a pairwise correlation satisfying the following conditions
\begin{align}
\Phi^\sigma_{\tau,\zeta}(\|\mathbf{r}_i-\mathbf{r}_j\|)= 
\left\{\begin{array}{cc}
0 & \mbox{ as } \|\mathbf{r}_i-\mathbf{r}_j\|\rightarrow 0, \\
1& \mbox{ as } \|\mathbf{r}_i-\mathbf{r}_j\|\rightarrow\infty.
\end{array}\right.
\end{align}

Commonly used radial basis functions include generalized exponential functions
\begin{align}
\Phi^{\rm E}_{\tau,\kappa} (\|\mathbf{r}_i-\mathbf{r}_j\|) = e^{-(\|\mathbf{r}_i-\mathbf{r}_j\|/\tau(r_i+r_j))^{\kappa}},\quad \kappa>0,
\end{align}
and generalized Lorentz functions
\begin{align}
\Phi^{\rm L}_{\tau,\nu} (\|\mathbf{r}_i-\mathbf{r}_j\|)= \dfrac{1}{1+(\|\mathbf{r}_i-\mathbf{r}_j\|/\tau(r_i+r_j))^{\nu}},\quad \nu>0,
\end{align}
where $r_i$ and $r_j$ are, respectively, the van der Waals radius of the $i^{th}$ and $j^{th}$ atoms.

Centrality is widely used in graph theory or network analysis to describe node  importance \cite{borgatti2005centrality}. 
Specifically,  closeness and harmonic centralities  are defined as $1/\sum_j \|\mathbf{r}_i-\mathbf{r}_j\|$  and  $\sum_j 1/\|\mathbf{r}_i-\mathbf{r}_j\|$, respectively.  
The degree of centrality simply counts  the number of edges upon a node. 
Our atomic centrality for $i^{th}$ atom can be regarded as  an extension of the harmonic formulation
\begin{eqnarray}
\mu^{k,\sigma,\tau,\nu,w}_i=\sum_{j=1}^{|V_{{\cal P}_k }|}w_{ij}\Phi^{\sigma}_{\tau,\nu}(\|\mathbf{r}_i-\mathbf{r}_j\|), \nonumber
\\
\quad \alpha_i\beta_j \in \mathcal{P}_k,
\quad \forall i=1,2,\ldots, |V_{{\cal P}_k}|,
\end{eqnarray}
where  $w_{ij}$ is a weight function assigned to each atomic pair, with $w_{ij}=1$ for atomic rigidity or $w_{ij}=q_j$ for atomic charge.   

In order to describe a centrality for the {whole} MWCS $G(V_{\mathcal{P}_k},E_{\mathcal{P}_k}^{\sigma,\tau,\zeta})$, we take into account a summation of the atomic centralities
\begin{align}
\mu^{k,\sigma,\tau,\nu,w} = \sum_{i=1}^{|V_{{\cal P}_k}|}\mu^{k,\sigma,\tau,\nu,w}_i.
\end{align}
It is this subgraph centrality that makes partition $\{ {\rm CN}\}$ equivalent to partition $\{ {\rm NC}\}$. 

Since we have 15 choices of the set of weighted colored edges ${\cal P}_k$, we can obtain corresponding 15 subgraph centralities $\mu^{k,\sigma,\tau,\nu,w}$. By varying kernel parameters $(\sigma,\tau,\nu,w)$, one can achieve multiscale centralities for multiscale weighted colored subgraph (MWCS) \cite{bramer2018multiscale}. For a two-scale WCS, we obtain a total of 60 descriptors for a protein.

Together with vertices $V$, the collection of all edges $E=\{		E_{\mathcal{P}_k}^{\sigma,\tau,\zeta} |  k=1,2, \dots,15\}$ defines  weighted graph  $G(V, E)$. However,  here $G(V, E)$ has a limited descriptive power in machine learning prediction. 
MWCSs  $G(V_{\mathcal{P}_k},E_{\mathcal{P}_k}^{\sigma,\tau,\zeta})$ and their centralities $\mu^{k,\sigma,\tau,\nu,w}$
are used in the present work to describe protein electrostatics.

\subsection*{Machine learning algorithms}\label{ML}
\subsubsection*{General description} 
In the present work, the prediction of PB electrostatic solvation free energy is formulated as a standard supervised learning. The training data set $\mathcal{D}$ can be expressed as 
  $$\mathcal{D}= \{ ({\bf x}^{(i)}, y^{(i) })  |  {\bf x}^{(i)}\in \mathbb{R}^n,   y^{(i) } \in \mathbb{R}, i=1,\cdots, M\},$$
where ${\bf x}^{(i)}$ is the feature vector for the $i$th sample in the training set, 
$ y^{(i) }=\Delta G^{(i)}$ as label is the electrostatic solvation free energy of the $i$th sample, 
$n$ and $M$ are the sizes of feature vector and the training set. 
$\Delta G^{(i)}$ will be given by the accurate MIBPB solver, which is justified by a convergence analysis in the Result section.  
The feature vector will be generated from the graph theory and the GB model.  

A variety of machine learning algorithms, 
including linear regression (LR), 
random forest (RF), 
gradient boosting decision tree (GBDT)  
and deep neural network (DNN) 
can be applied to predict the electrostatic free energy of the PB model. 
LR is a simple approach designed for the linear approximation of the mapping. 
RF and GBDT are  both decision tree based ensemble methods. 
RF builds a large number of uncorrelated trees and utilizes bootstrap and aggregating (i.e., bagging).  
GBDT makes use of gradient descent in conjugation with the boosting procedure, 
which successively introduces weak learners to compensate for the errors of existing learners. 
DNN methods become powerful when errors are back-propagated to correct neural weights. 
However, DNN methods typically involve a large number of weights and thus are subject to overfitting.  
DNN methods might not offer better predictions unless the size of the training data is sufficiently large. 

\subsubsection*{Feature Descriptions}
Our ML model currently uses \underline{367} features considering protein structures, force field, graph theory representation, etc. 
\begin{itemize}
\item GB model related features (\underline{240}) : For each of the 15 paired elements among \{C,N,S,O,H\}, there are 16 features as  
GB terms, absolute GB terms, intermediate GB terms, absolute intermediate GB terms, charges with dielectric, absolute charges with dielectric, GB charge terms, absolute GB charge terms, charges, absolute charges, plus 6 rigidity indeces. 
We use the BornRadius code \cite{Onufriev:2000} developed in Onufriev's group wrapped with python scripts to generate these features. 
\item Protein features (\underline{51}):
area (7) (C, N, O, S, H, CNOS, CNOSH), 
charge (7), 
absolute charge (7), 
van der Waals force ($15$), 
Coulomb force ($15$). 
\item  Environment features (\underline{31}): volume (6), hydropathy (6), area (6), weight (6), pharma (6), sum of charge (1)
 \item Features from graph theory representation (\underline{45}): Exp and Lor kernels ($15\times3$ kernels)
\end{itemize}

In this work, we use MWCS centrality $\mu^{k,\sigma,\tau,\nu,w}$ to generate a mathematical  representation of biomolecular electrostatics. 
In general, we seek an intrinsically low-dimensional representation of electrostatic solvation free energies for a large set of proteins.  
Specifically, we consider  both the  exponential kernel ($\sigma={\rm E}$) and the Lorentz kernel ($\sigma={\rm L}$). 
Each kernel is parametrized  with both the weight selections of rigidity ($w=1$) and charge ($w=q$). 
We select the powers of the exponential kernel $\nu=2$ and the Lorentz kernel $\nu=3$. 
The atomic features with 15 types of element partitions ($\mathcal{P}_k$, $k=1,2,\dots,15$) are considered. 
However, the scale distance factor $\tau$ is to be optimized during the training and testing processing. Additionally, the GB electrostatic solvation free energies is employed as a special feature that incorporated seamlessly with machine learning algorithms.   

\subsubsection*{Generalized Born (GB) methods based gradient boosting decision tree (GBDT)}\label{sub_GBbasedGBDT}
The main idea GBDT model is to first use the features and the labels to build a decision tree model, which is able to give predicted labels. Then the residue between the original labels and the predicted labels will be used as new labels, together with the original features to build another decision tree model, which gives another predicted labels. 
Using the difference between the initial labels and the predicted labels as the new labels, 
this procedure can be done recursively and final predicted labels will be the summation of all predicted labels.    
The model is optimized by minimizing the cost function between the initial labels and their predicted values using the gradient descent method.  

In our framework of GB-based GBDT model, our first decision tree is the GB model, 
whose solvation energy is treated as the first predicted labels to initial labels, 
the PB-model based solvation energy.  
The rest of trees are built from our MWCS features. 
The loss function depends on a number of trees, 
the structure of trees  and MWCS features. 
In this work, the loss function $L$ will also be optimized with respect to MWCS parameters. 
The details of this model can be found in the supplementary material.

\subsubsection*{Generalized Born (GB) methods based deep neural network (DNN)}\label{GBbasedGBDNN}
For DNN, we use the 367 features for each protein as the inputs to the network for training, test, and prediction purposes. In our model, the label is defined as $y^{(i)} =\Delta G_i^{\rm PB} - \Delta G_i^{\rm GB}$ for $i=1,\cdots, M$, where $\Delta G_i^{\rm GB}$ is calculated as the core feature used outside the network. This core feature gives the global estimate while other features fed into the network provides local details. The quantity $\Delta G_i^{\rm PB}$ is obtained as training/test data by solving the PB equation with MIBPB at refined mesh (e.g. h=0.2). In prediction, we have $\Delta \hat{G}_i^{\rm PB}=\Delta G_i^{\rm GB} + \hat{y}^{(i)}$, where
 $\hat{y}^{(i)}$ is the predicted value from the DNN. The DNN has multiple layers and the weights of the network are obtained by back-propagation. We tune the parameters of the network by sampling the parameter space to receive an optimized combination of the parameters for best prediction accuracy. 

\section*{Results}

\label{ResultsDiscussion}
This section reports results from the proposed PBML model.  
Evaluation metrics, data selection, machine learning label calculation, and feature selection are discussed before results are shown. 
We first justify the choice of MIBPB solver \cite{DuanChen:2011a} to generate labels 
compared with popular PB solvers such as Amber \cite{PBSA:2009} and DelPhi \cite{Delphi:2012}.
We demonstrate the accuracy and efficiency of the proposed PBML model in electrostatic solvation free energy predictions.   
The numerical results of MIBPB, Amber and DelPhi are generated with 
an Intel Xeon E5-2670v2 from HPCC of Michigan State University (MSU), and the machine learning results are produced on a desktop with Intel Core I5 7500 and 16G Memory, 
using the scikit-learn python package. 
The electrostatic solvation free energies are generated 
with the room temperature $T=298.15\text{K}$ and dielectric constant $\epsilon_1 = 1$, and $\epsilon_2=80$.

\subsection*{Evaluation metrics}

Through this paper, we use the mean absolute percentage error (MAPE) and absolute relative error (ARE)  for the analysis of prediction accuracy, which are defined as 
\begin{equation}\label{EMAPE} 
E_{\rm MAPE}= \frac{100\%}{M}\sum^{M}_{i=1}\Big|\frac{y^{(i)}-\hat{y}^{(i)}}{y^{(i)}}\Big|, E_{\rm ARE}= \Big|\frac{y^{(i)}-\hat{y}^{(i)}}{y^{(i)}}\Big|.\nonumber
\end{equation}
where $y^{(i)}$ is the $i$th label, i.e., the PB electrostatic solvation free energy of the $i$th molecule and $\hat{y}^{(i)}$ is the predicted value. 
%

\subsection*{Data preparation}

In the present work, 
the selected 4294 protein structures are obtained from the PDBbind v2015 refined set and PDBbind v2018 refined set as the training set \cite{PDBBind:2015}. The PDBbind v2015 core set of 195 proteins as listed in Supplementary Material is adopted as the test set.  The training set has proteins sized  from 997 to 27,713 atoms while the test set proteins range from 1,702 to 26,236 atoms. 

A data pre-processing is required before a PB solver can be called. 
The protein structures in the original data set are protein-ligand complexes. 
Missing atoms and side-chains are filled using the protein preparation wizard utility of the Schrodinger 2015-2 Suite with default parameter setting. 
The Amber ff14SB general force field is applied for the atomic van der Waals radii and partial charges.

\subsection*{Simulation Results}
\subsubsection*{Convergence comparison of the PB solvers}
\begin{figure}[h!]
\begin{tabular}{cc}
\raisebox{105pt}{(a) 195 proteins }&~\includegraphics[width=2.2in]{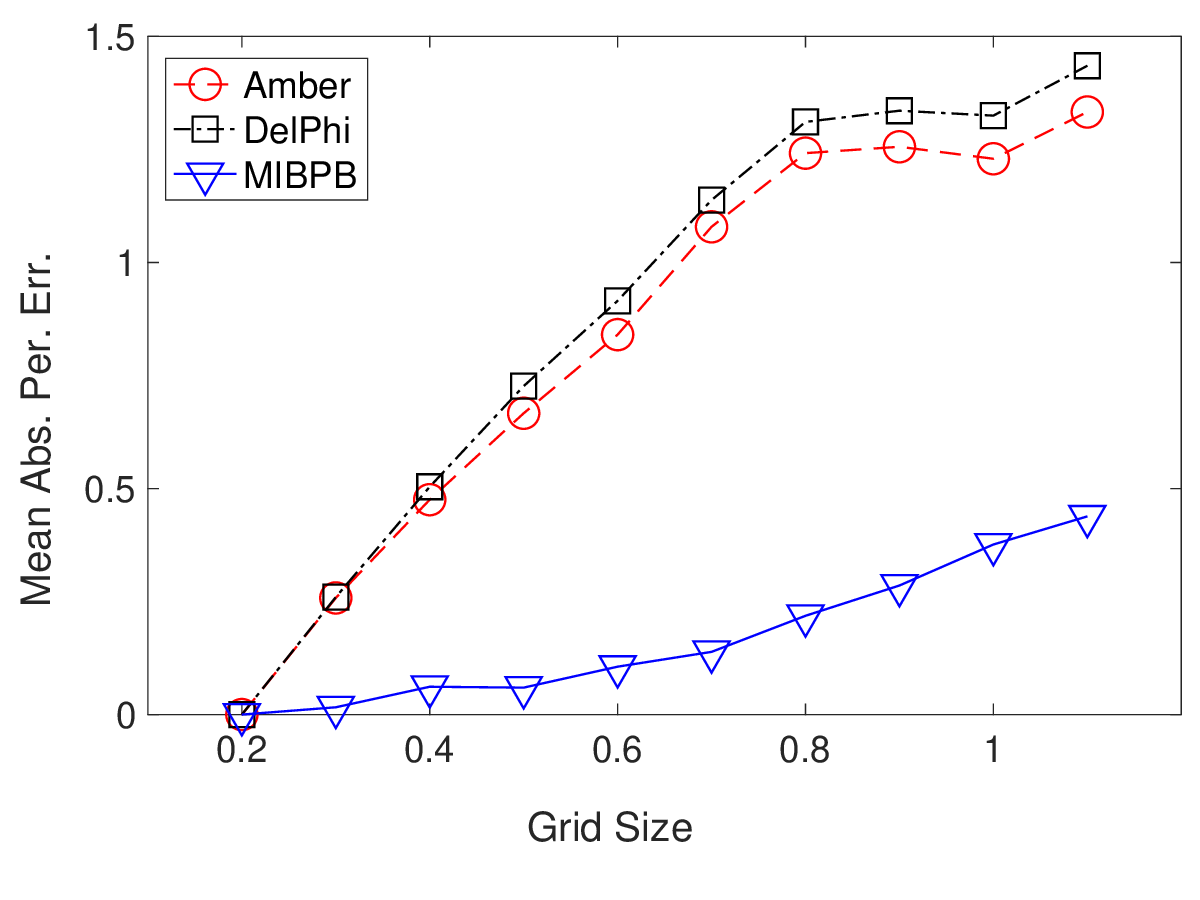} \\
\raisebox{75pt}{(b) protein 3gnw}&\includegraphics[width=2.2in]{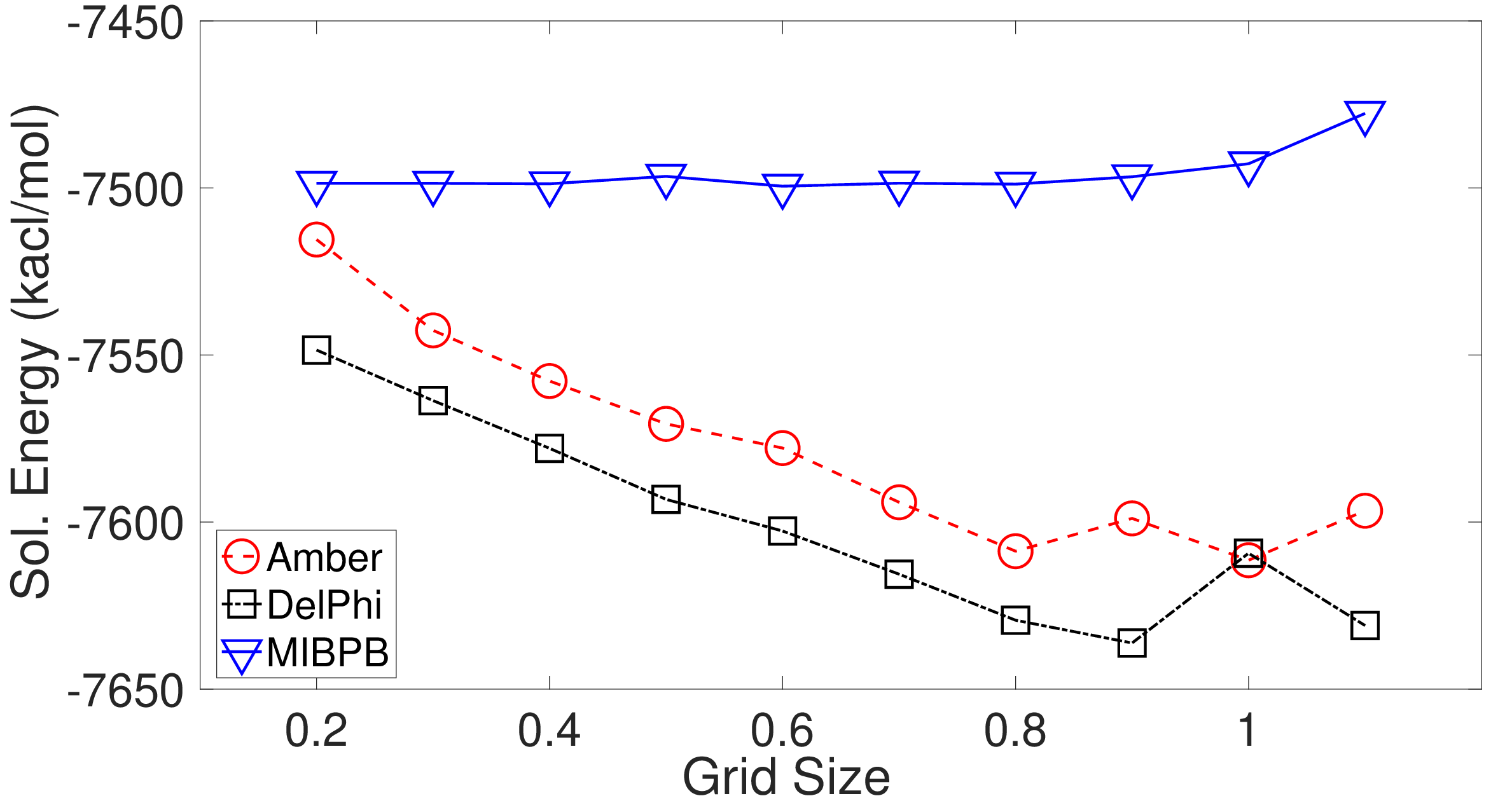} \\
\raisebox{75pt}{(c) protein 3owj} &\includegraphics[width=2.2in]{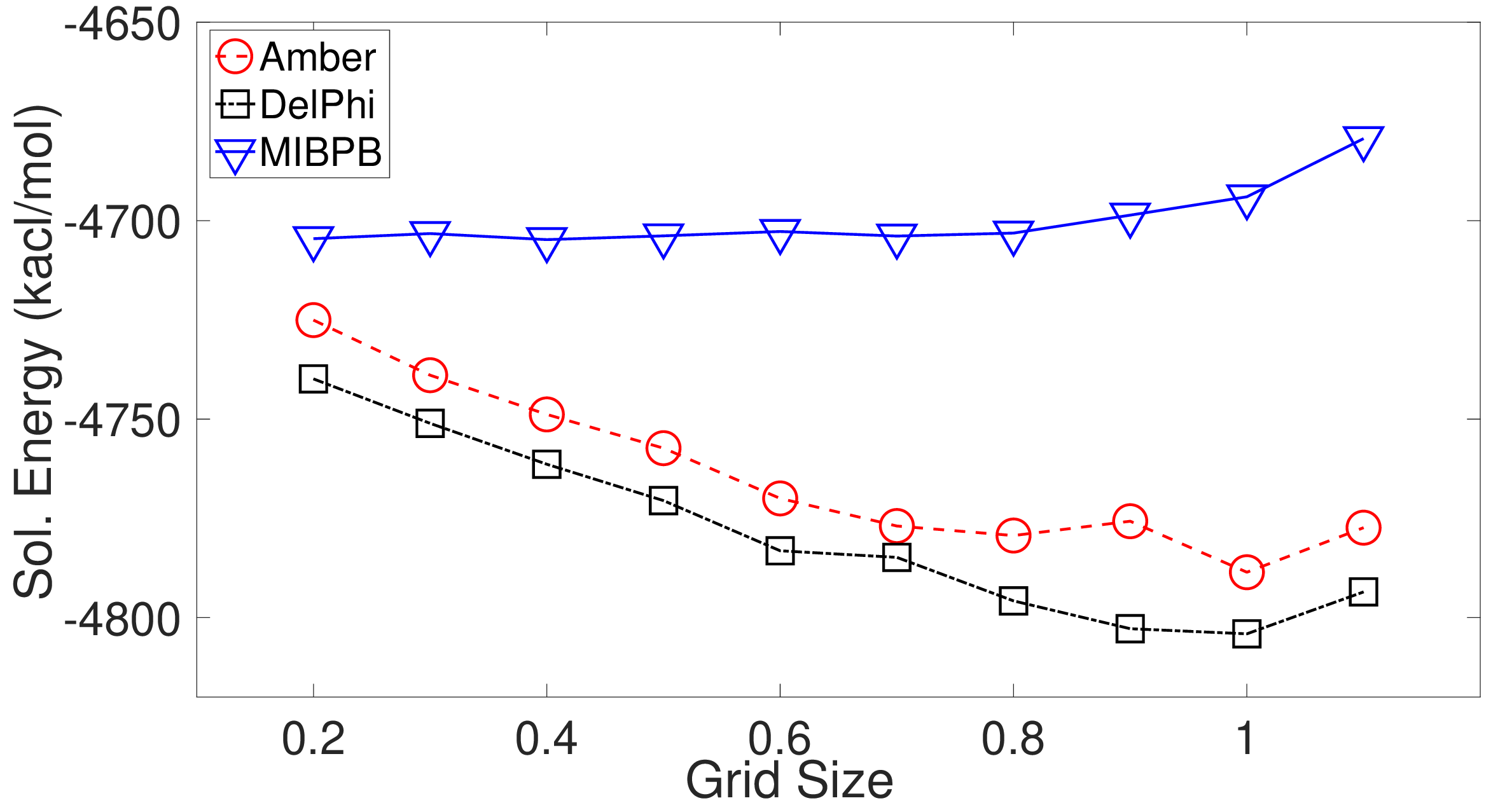}\\
\end{tabular}
\caption{\small Convergence comparison among Amber, DelPhi, and MIBPB; (a) MAPEs at ten grid sizes for Amber, DelPhi, and MIBPB in computing the electrostatic solvation free energies of  195 test proteins. 
For a protein in each method, the reference value is computed at the mesh  size of  0.2 \AA; 
(b)-(c) Illustration of the electrostatic solvation free energies obtained by Amber, DelPhi, and MIBPB at ten different mesh sizes from 0.2 to 1.1\AA ~for proteins 3gnw and 3owj. }
\label{fgr:self_convergence}
\end{figure}

We first carry out the convergence analysis of three PB solvers for the test set of 195 proteins 
to justify the use of MIBPB solver to produce accurate electrostatic solvation energy as the labels. 
For each protein,  we compute their electrostatic solvation free energies at ten different mesh sizes, 
ranging from 0.2 \AA ~ to 1.1 \AA.  
For each PB solver, its results  at finest mesh size 0.2 \AA~ are used as the references 
to evaluate the relative errors for other meshes.  
As shown in Fig.~\ref{fgr:self_convergence}(a), the MAPEs using Amber and DelPhi are less than 1.5 \%, but that from MIBPB is less than 0.5\% at all mesh sizes. 
We next examine the electrostatic solvation free energies computed by three PB solvers  on two sampled proteins.  
As shown in Fig.~\ref{fgr:self_convergence}(b-c), using the test proteins 3gnw(b) and 3owj(c),  
the energies obtained by MIBPB do not change much over the mesh refinement, 
while those computed by Amber and DelPhi vary more significantly. 
We also observed that energies obtained by Amber and DelPhi  
converge toward those of MIBPB as the mesh is refined.  
These tests justify that MIBPB as the most accurate method among these three PB solvers, to be used to compute labels for the ML models. Some further convergence tests and comparison between PB solvers can be found in supplementary material.

 \subsubsection*{Comparison between different ML models}
 After justifying the use of MIBPB solver to generate the labels, 
 we next apply LR, RF, GBDT, and DNN to produce corresponding learned models using the training data set. 
 We then use these learned models to predict the solvation energy for the 195 proteins in the test set. 
 The MAPE for each learned model is shown in Table~\ref{tb:four_methods}. 
 The result shows that DNN has better performance than other three methods, 
 thus we use DNN as our ML algorithm for a further comprehensive training and test of the PBML model. 

\begin{table}[htp]
\caption{\small The MAPEs of LR, RF, GBDT, DNN for the test set of 195 proteins. For LR and RF, we use the default parameters. For GBDT, we set the learning rate 0.05, the number of estimators 1500,  maximum depth 5; 
The DNN is trained with 448 different combination of parameters and the final optimized choice has batch size 400, epoch 500, layers 367, 2048, 2048, 512, 512, 1.}
\begin{center}
\begin{tabular}{c|c|c|c|c}
\hline
& {LR} & {RF} & {GBDT} & {{\bf DNN}}\\\hline
MAPE & 1.5317 & 0.8238 & 0.4432 & {0.3796}\\
\hline
\end{tabular}
\end{center}
\label{tb:four_methods}
\end{table}%

\begin{figure}[h]
\includegraphics[width=3.3in]{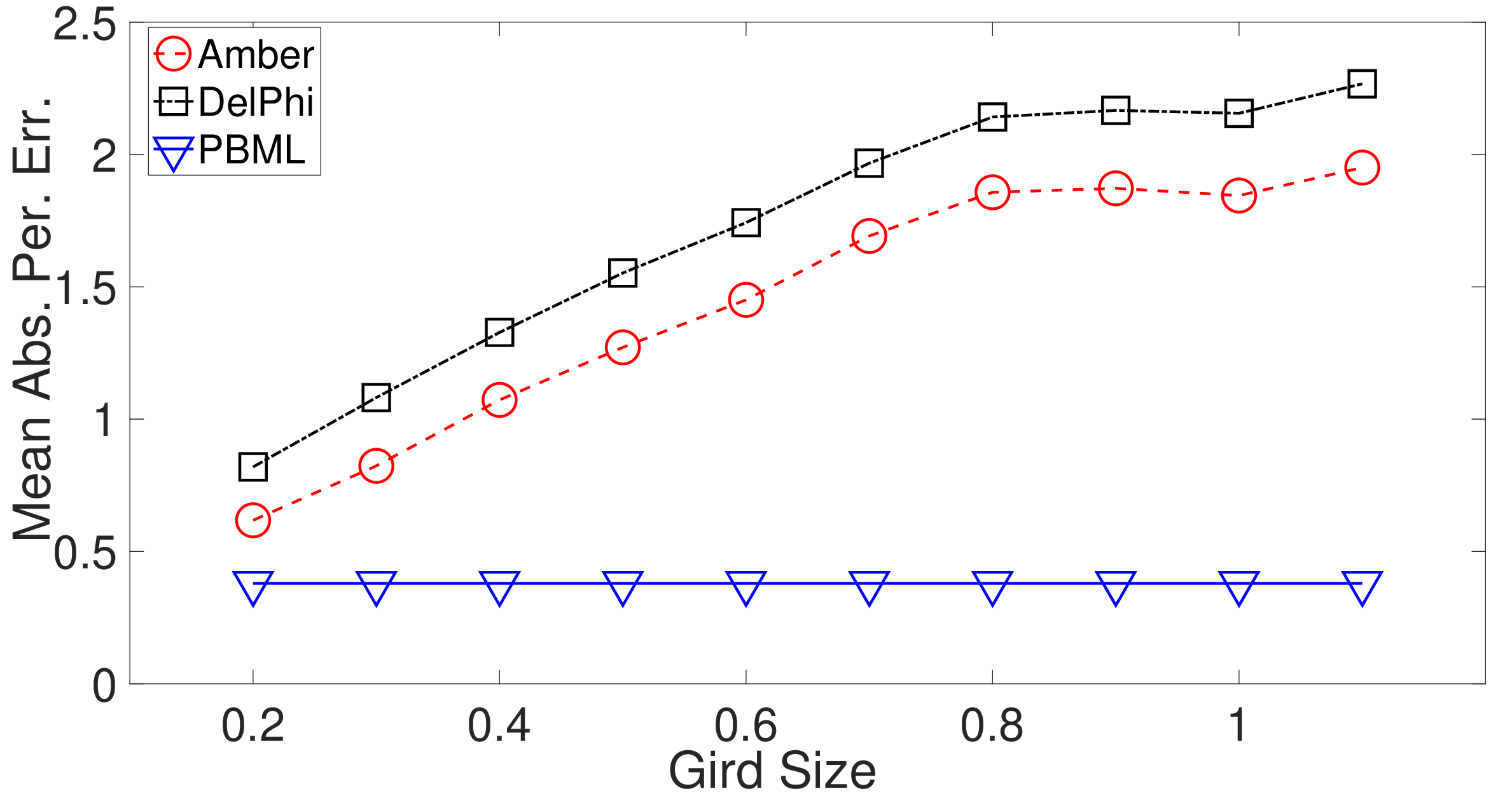}
\caption{\small 
Comparison of the MAPEs of Amber, DelPhi and PBML (use result in Table~\ref{tb:four_methods} from DNN model ) of the electrostatic solvation free energies of the test set at ten mesh sizes.  The reference values are the results of MIBPB at the grid size of 0.2 \AA. }
	\label{fgr:FeatureImportance}
\end{figure}

\subsubsection*{Performance of the PBML model}

Our final PBML model is essentially the GB-based DNN model, which uses the GB core feature with additional 367 features as described before. 
To understand the advantages of the model and its prediction, 
we plotted a comparison of the its MAPE with those of  Amber and DelPhi at ten mesh sizes in Fig.~\ref{fgr:FeatureImportance}. 
Note that although Amber and DelPhi MAPEs reduce significantly as the mesh is refined, 
even at the finest mesh of 0.2~\AA, these two methods have not reached the accuracy of PBML, which does not depend on grid size once the model is trained/learned.

\begin{figure}[h]
\begin{tabular}{cc}
\raisebox{115pt}{(a)}& \hskip -12pt \includegraphics[width=3.15in]{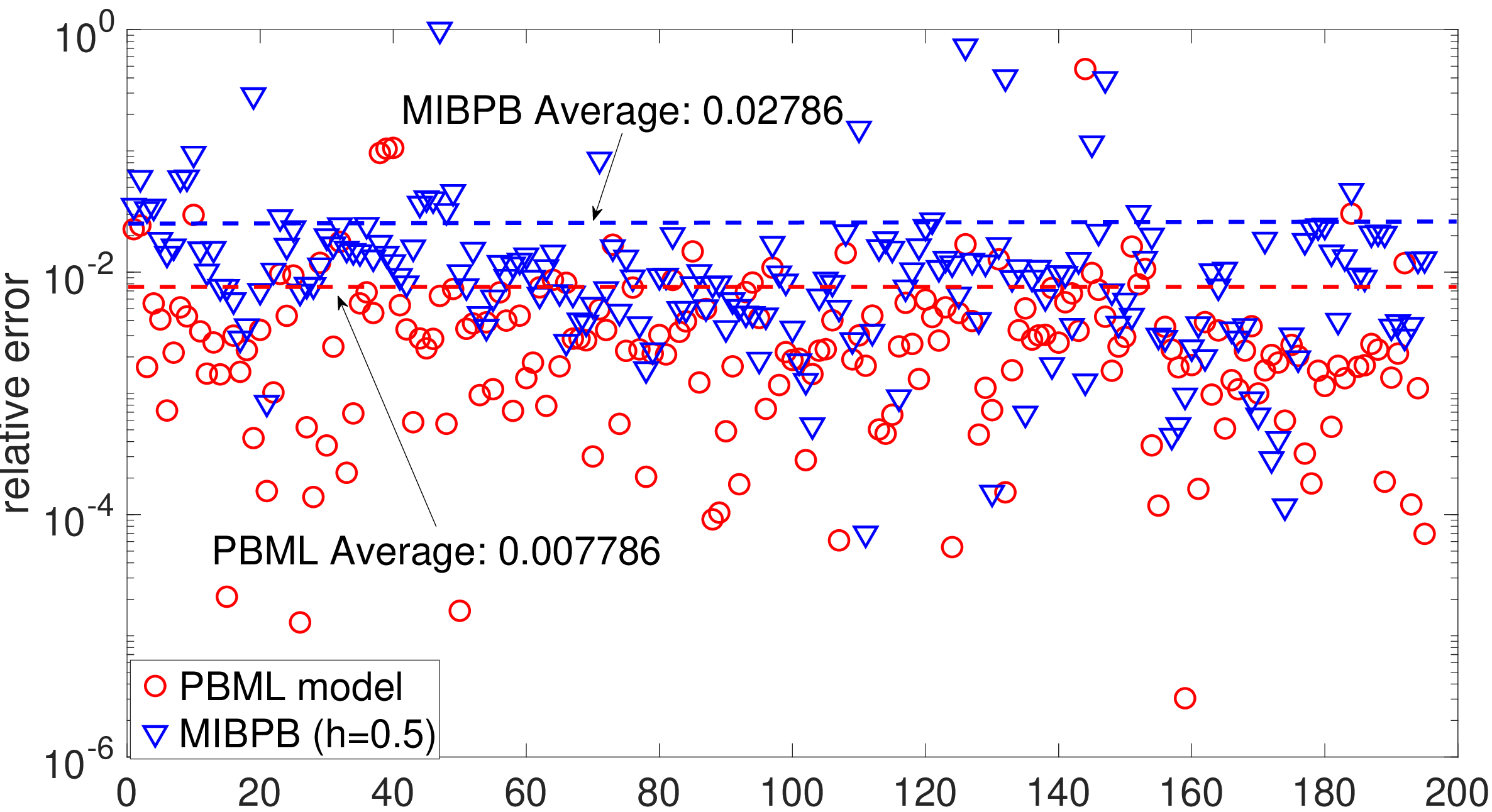}\\
\raisebox{115pt}{(b)}& \hskip -10pt  \includegraphics[width=3.1in]{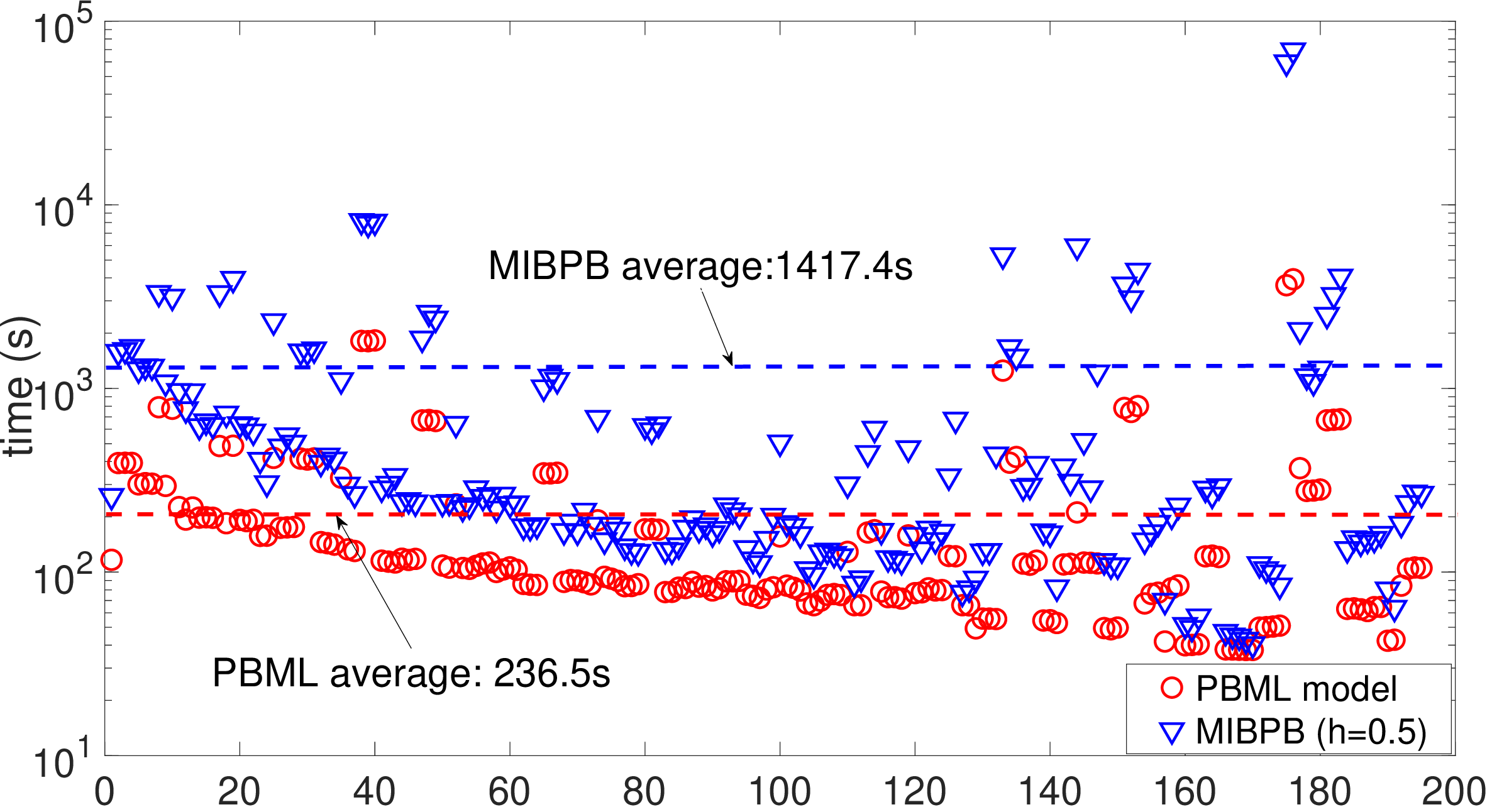}
\end{tabular}
\caption{\small Accuracy and efficiency comparison on computing solvation energy on 195 proteins using MIBPB at $h=0.5$ and DNN based PBML model:(a): relative error in solvation energy; (b): time.  The average relative errors for PBML and MIBPB are 0.00776 and 0.02786. The average time for PBML and MIBPB is 236.5s and 1417.4s respectively. 
}
\label{fgr:Comparison}
\end{figure}

To further check the accuracy and efficiency of our PBML model. We compute the solvation energy of 195 test proteins using both the MIBPB at $h=0.5$ and the PBML model. Note the PBML model is trained with the 4000+ protein training set as described before labeled by solvation energy computed using MIBPB at $h=0.2$. 
For this test, we use results from MIBPB at $h=0.2$ as benchmark values while use results from MIBPB at $h=0.5$ for the comparison with results from the PBML model for the 195 proteins in the test set. 
Fig.~\ref{fgr:Comparison}(a) shows the relative error in solvation energy and from individual sample or average we see the PBML model is obviously more accurate than the MIBPB model at $h=0.5$. 
Fig.~\ref{fgr:Comparison}(b) shows the elapsed time and from individual sample or average we see the PBML model is significantly more efficient than the MIBPB model at $h=0.5$. All figures are plotted using log scale in error and time since results from different proteins are very variant.

\section*{Conclusion}
This work introduces 
the Poisson-Boltzmann based machine learning (PBML) model for the prediction of  electrostatic solvation free energies of biomolecules. 
Our goal is to offer an efficient ML-based electrostatic analysis of new molecules or new conformations of molecular dynamics at a small fraction of time used in solving the Poisson-Boltzmann (PB) equation at a similar level of accuracy, or at a similar level of computational time but with a much higher accuracy than a commonly used PB solver can ever deliver. 
To this end, we first search the most accurate PB solver for generating ML labels. 
The second-order accurate MIBPB solver turns out to converge faster than  two other eminent  PB solvers, 
namely the DelPhi and the Amber PB solver. 
Additionally, we adopt multiscale weighted colored subgraph (MWCS) for ML feature generations, 
which produces excellent low-dimensional intrinsic representations of biomolecules.
One global core feature is computed from the generalized Born (GB) model. 
To maintain the efficiency, we employ a few ML algorithms, including linear regression (LR), random forest (RF), gradient boosting decision tree (GBDT), and Deep Neural Network (DNN). 
It is found that the present PBML model using DNN can more efficiently and accurately produce electrostatics over traditional grid-based PB solvers. 


\section*{Author Contributions}

JC, CZ, WG, and GWW designed the research. JC, YX, and XY carried out all simulations, analyzed the data. JC, WG, and GWW wrote the article.

\section*{Acknowledgments}
The work of WG is supported in part by NSF grant DMS-2110922.
The work GWW is supported in part by NSF grants DMS-2052983 and IIS-1900473 and  NIH grants R01AI164266, and R35GM148196. 

\section*{Literature cited}
\renewcommand\refname{}

\bibliography{refs}


\clearpage
\section*{Supplementary Material}
An online supplement to this article can be found by visiting BJ Online at \url{http://www.biophysj.org}.

\subsection*{Generalize Born (GB) methods based gradient boosting decision tree (GBDT)}\label{GBbasedGBDT}
It is noted that GB model provides a good approximation to PB model. As such, its  $\Delta G^{\rm GB}$ can be incorporated   into the GBDT algorithm. 
To this end, we outline  the GB method based GBDT algorithm as follows. 
Normally, the first step is to build a decision tree ${\rm T}_1$ to fit $\{ ({\bf x}^{(i)}, y^{(i) }) \}_{i=1}^M$, leading to  predicted labels  $\{ p_1({\bf x}^{(i)})  \}_{i=1}^M$. 
The errors (or residues) of the predictions are $r_2^{(i)}=y^{(i)}-p_1({\bf x}^{(i)})$. 
If  {$r_2^{(i)}\neq 0$} for some $i$, 
one builds another decision tree  ${\rm T}_2$ to fit $\{ ({\bf x}^{(i)}, r_2^{(i) }) \}_{i=1}^M$, 
leading to new predicted labels  $\{ p_2({\bf x}^{(i)}  \}_{i=1}^M$. 
The errors are  $r_3^{(i)}=r_2^{(i)}-p_2({\bf x}^{(i)})=  y^{(i)}-p_1({\bf x}^{(i)}) -p_2({\bf x}^{(i)})$. The predicted labels for  $\{ ({\bf x}^{(i)}, y^{(i) }) \}_{i=1}^M$ are { $\{ p_1({\bf x}^{(i)})+ p_2({\bf x}^{(i)})  \}_{i=1}^M$}. 
If $|r_3^{(i)}|>0$, one can build ${\rm T}_3$ to fit $\{ ({\bf x}^{(i)}, r_3^{(i) }) \}_{i=1}^M$, leading to  new predicted labels   $\{ p_3({\bf x}^{(i)})  \}_{i=1}^M$. 
In general, the predicted model for the data $\{ ({\bf x}^{(i)}, y^{(i) }) \}_{i=1}^M$ based on $K$ consecutive decision trees is 
\begin{equation}
 \hat{y}_{K}^{(i)}=\sum_{k=1}^{K}p_k({\bf x}^{(i)}), i=1,2,\cdots,M. 
\end{equation}
This is the so called boosting tree procedure. One can setup a loss function 
\begin{equation}\label{Loss}
 L_k= \sum_{i=1}^M l_k\left(  y^{(i)}-\hat{y}_{k}^{(i)}\right)=\sum_{i=1}^M\frac{1}{2} \left(  y^{(i)}-\hat{y}_{k}^{(i)}\right)^2,
\end{equation}
 to minimize the loss via the gradient descent optimization of decision trees. A general minimization  procedure is the follows. 
For $k=2$ to $K$, 
\begin{itemize}
\item Calculate gradient 
\begin{equation}\label{Loss2}
r_k^{(i)}=-\frac{\partial l_{k-1}}{\partial p_{k-1}({\bf x}^{(i)})}
\end{equation}
\item Construct  decision tree ${\rm T}_{k}$ to fit  $\{ ({\bf x}^{(i)}, r_k^{(i) }) \}_{i=1}^M$, leading to $ p_{k} $ as the learner function of ${\rm T}_{k}$.  

\item Choose learning rate $\alpha_k$ such that 
\begin{equation}\label{Loss3}
\alpha_k := {\rm argmin}_{\alpha} \sum_{i=1}^M l_{k-1}\left(  y^{(i)}-\hat{y}_{k-1}^{(i)}\right)+ \alpha p_k({\bf x}^{(i)})
\end{equation}

\item Update model 
\begin{equation}\label{Loss4}
 \hat{y}_{k}^{(i)}=\hat{y}_{k-1}^{(i)}+\eta \alpha_k p_k,
\end{equation}
where $\eta$ is the shrinkage, a predefined parameter.  
 \end{itemize}
In our work, we select  ``${\rm T}_1$'' to be the GB model which leads to $\{ {p_1({\bf x}^{(i)})}=\Delta G_i^{\rm GB} \}_{i=1}^M$. The rest of trees are built from our MWCS features. The loss function depends on a number of trees, the structure of trees  and MWCS features. In this work, the loss function $L$ will also be optimized with respect to MWCS parameters.

\subsection*{Convergence comparison between Amber, Delphi, and MIBPB on 30 selected proteins}

\begin{figure}[h!]
\centering
\includegraphics[height=2.5in]{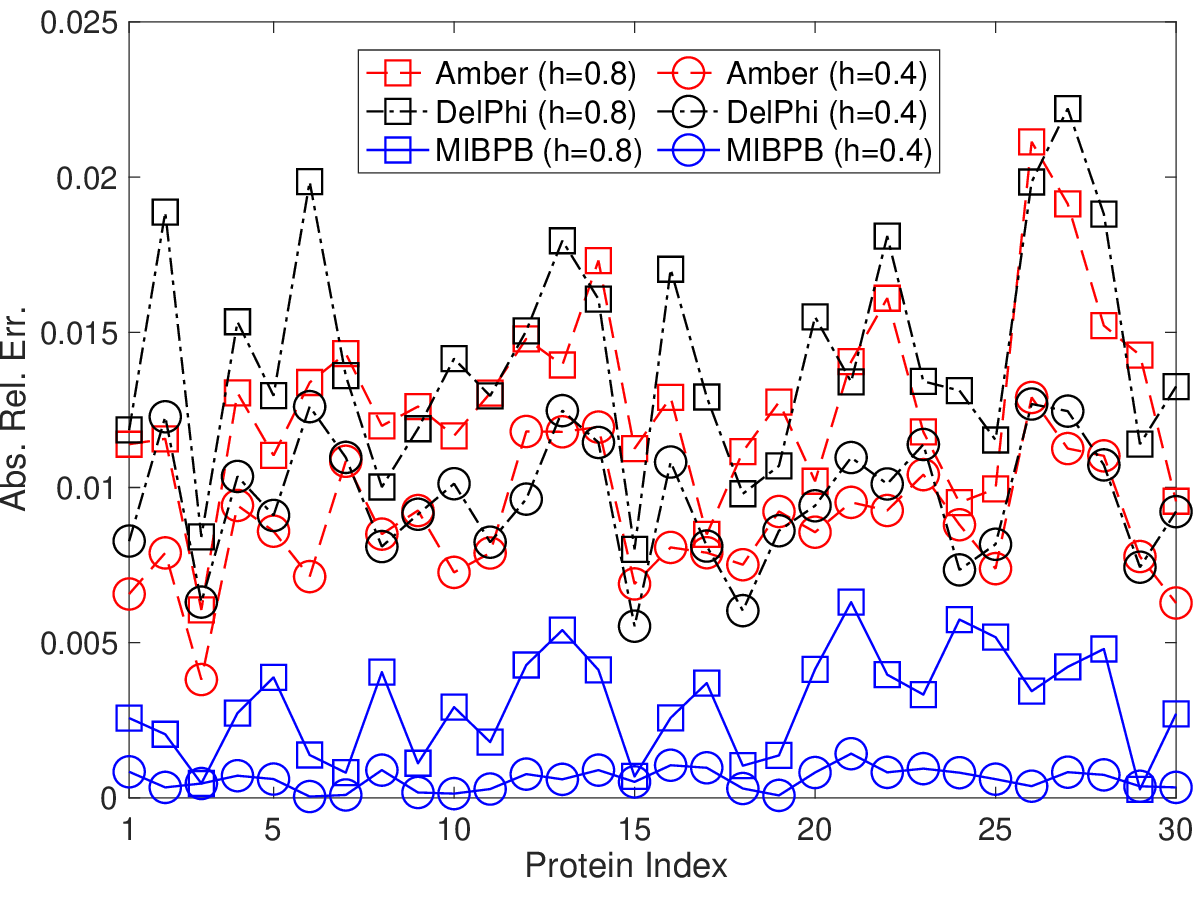}
\caption{\small Convergence comparison among Amber, DelPhi, and MIBPB. 
The graph shows Absolute relative errors of Amber (dashed lines), DelPhi (dash-dot lines),  and MIBPB (solid lines) at mesh sizes 0.4 \AA~ (cycles)  and 0.8 \AA~(squares)  for 30 proteins. For a protein in each method,  the reference value is calculated at the mesh  size of  0.2 \AA.}
\end{figure}
To further investigate the convergences of the three PB solvers, 
we randomly picked 30 proteins from the core set  
and plot their AREs at two mesh sizes 0.4 \AA~(circle)  and 0.8 \AA~(square) compared with results at 0.2 \AA~in Fig. \ref{fgr:self_convergence}(b). We see that Amber and DelPhi offer a similar level of convergence, 
while MIBPB returns significantly smaller errors. These 30 proteins are:
2qmj,
3ebp,
2x8z,
3bkk,
1gpk,
1f8b,
3ge7,
3huc,
3pe2,
3mfv,
2qbr,
3imc,
1saq,
3ivg,
3b3w,
3mss,
2v7a,
2zcq,
3utu,
3u9q,
3kwa,
3gbb,
1uto,
2yge,
2iwx,
3bpc,
2pcp,
2gss,
4dew,
3nox.

\subsection*{The list of 195 proteins from testing set}
Below is the list of the 195 proteins used for testing purpose. \\
1lor,
1ps3,
3d4z,
3ejr,
2qmj,
3l4w,
3l4u,
3l7b,
3g2n,
3ebp,
2w66,
2wca,
2vvn,
2x97,
2xhm,
2x8z,
2x0y,
2cbj,
2j62,
3bkk,
3l3n,
2xy9,
1gpk,
1h23,
1e66,
3cj2,
2d3u,
3gnw,
3f3a,
3f3c,
3f3e,
4gqq,
1u33,
1xd0,
2wbg,
2j78,
2cet,
2zxd,
2zwz,
2zx6,
3udh,
4djv,
4gid,
3fk1,
2qft,
2pq9,
1f8d,
1f8b,
1f8c,
1n2v,
1r5y,
3ge7,
3huc,
3gcs,
3e93,
1q8t,
1q8u,
3ag9,
3owj,
2zjw,
3pe2,
2v00,
3pww,
3uri,
3mfv,
3f80,
3kv2,
2hb1,
2qbr,
2qbp,
3fcq,
1os0,
4tmn,
3pxf,
2xnb,
2fvd,
1qi0,
1w3k,
1w3l,
3imc,
3ivg,
3coy,
3b3s,
3b3w,
3vh9,
3mss,
3k5v,
2v7a,
2brb,
3jvs,
1nvq,
3acw,
2zcr,
2zcq,
1bcu,
1oyt,
3utu,
3u9q,
2yfe,
2p4y,
3uo4,
2wtv,
3myg,
3kgp,
1o5b,
1sqa,
3kwa,
2weg,
3dd0,
2xdl,
1yc1,
2yki,
1p1q,
3bfu,
4g8m,
3g2z,
4de2,
4de1,
1vso,
3gbb,
3fv1,
2y5h,
2xbv,
1mq6,
1loq,
1lol,
1uto,
3gy4,
1o3f,
2yge,
2iwx,
2vw5,
2ymd,
2xys,
2x00,
2r23,
3bpc,
1kel,
3ozt,
3oe5,
3nw9,
1zea,
2pcp,
1igj,
1lbk,
2gss,
10gs,
3su5,
3su2,
3su3,
3n7a,
3n86,
2xb8,
3ao4,
3zsx,
3zso,
3nq3,
3ueu,
3uex,
3lka,
3ehy,
3f17,
3cft,
4des,
4dew,
3dxg,
1w4o,
1u1b,
3ov1,
3s8o,
1jyq,
1a30,
3cyx,
4djr,
3i3b,
3muz,
3vd4,
2vo5,
2vl4,
2vot,
1n1m,
2ole,
3nox,
1hnn,
2g70,
2obf,
1z95,
3b68,
3g0w,
1sln,
2d1o,
1hfs,
2jdy,
2jdm,
2jdu.

\end{document}